\documentclass[%
 reprint,
 amsmath,amssymb,
prl,
]{revtex4-1}

\usepackage{graphicx}
\usepackage{dcolumn}
\usepackage{bm}
\usepackage{epstopdf}
\usepackage{color}

\begin{document}
\newcommand{\todo}[2]{\textcolor{red}{\textsc{#1}}}
\preprint{APS/123-QED}
\title{Narrowband quantum emitters over large spectral range with Fourier-limited linewidth in hexagonal boron nitride} 

\author{A. Dietrich$^1$, M. B\"urk$^1$, E. S. Steiger$^1$, L. Antoniuk$^1$, T. T. Tran$^2$, M. Nguyen$^2$, I. Aharonovich$^2$, F. Jelezko$^{1,3}$, A. Kubanek$^{1,3}$}
 \affiliation{$^1$ Institute for Quantum Optics, Ulm University,D-89081 Ulm, Germany\\
 $^2$School of Mathematical and Physical Sciences, University of Technology Sydney, Ultimo, New South Wales 2007, Australia\\
 $^3$ Center for Integrated Quantum Science and Technology (IQst), Ulm University,D-89081 Ulm, Germany}

\date{\today}

\begin{abstract}

Single defect centers in layered hexagonal boron nitride (hBN) are promising candidates as single photon sources for quantum optics and nanophotonics applications. However, until today spectral instability hinders many applications. Here, we perform resonant excitation measurements and observe Fourier-Transform limited (FL) linewidths down to $\approx 50$ MHz. We investigate optical properties of more than 600 quantum emitters (QE) in hBN. The QEs exhibit narrow zero-phonon lines (ZPL) distributed over a spectral range from 580 nm to 800 nm and with dipole-like emission with high polarization contrast. The emission frequencies can be divided into four main regions indicating distinct families or crystallographic structures of the QEs, in accord with ab-initio calculations. 
Finally, the emitters withstand transfer to a foreign photonic platform - namely a silver mirror, which makes them compatible with photonic devices such as optical resonators and paves the way to quantum photonics applications including quantum commmunications and quantum repeaters.



\end{abstract}

\maketitle


QE with spectral linewidth at the FL are the building blocks for various quantum information applications such as quantum repeaters \cite{vuckovic_design_2001,senellart_high-performance_2017} or quantum networks \cite{kimble_quantum_2008}.
Operating the QE at the FL is essential since spectral diffusion, dephasing or competing phonon-processes hinder quantum applications that rely on an efficient spin-photon interface \cite{moehring_entanglement_2007,ritter_elementary_2012,bernien_heralded_2013} or indistinguishable, single photons \cite{legero_quantum_2004,lettow_quantum_2010,sipahigil_quantum_2012}.
Single-photon sources with FL lines have been demonstrated with atoms or ions in gas phase, organic molecules as well as with isolated solid-state quantum systems at cryogenic temperatures such as color centers or quantum dots \cite{lounis_single-photon_2005,senellart_high-performance_2017}. 
\\
A new emerging class of QEs are embedded in two-dimensional host materials such as hBN, that offer single photon emission covering the complete optical spectrum and enable novel architectures for integrated quantum technology \cite{schell_coupling_2017,tran_deterministic_2017}.
The optical properties are promising with high photostability \cite{kianinia_robust_2017}, high brightness \cite{tran_quantum_2016_nature,martinez_efficient_2016,grosso_tunable_2017}, large Debye Waller factor \cite{tran_robust_2016,li_nonmagnetic_2017}, good polarization contrast  \cite{jungwirth_temperature_2016,chejanovsky_structural_2016,exarhos_optical_2017}, spectral tunability \cite{grosso_tunable_2017,tran_robust_2016}, stable optical lines as narrow as 45 $\mu$eV at cryogenic temperatures \cite{li_nonmagnetic_2017,jungwirth_optical_2017} and, very recently, resonant excitation with lines measured in photoluminescence excitation (PLE) as narrow as 1 GHz \cite{tran_resonant_2017}.
However, the close proximity of the QE to the host surface makes it susceptible to spectral instability, typically exhibiting, blinking, rapid spectral diffusion and pure dephasing rates much larger than the population decay rates and therefore ZPLs much broader than the natural linewidth \cite{sontheimer_photodynamics_2017}.
\\
In this work, we demonstrate spectrally stable, single-photon emission from defect center in hBN under resonant excitation without significant spectral diffusion for as long as 30 seconds. 
We perform a detailed study of the emission properties of 627 different defect centers in hBN with central emission frequencies from 580 nm to 800 nm in  to previous measurements \cite{tran_resonant_2017,jungwirth_temperature_2016,chejanovsky_structural_2016,tran_quantum_2016,bourrellier_bright_2016}.
Clustering of the emission frequencies into four main groups delivers a new experimental reference for refined ab-initio simulations with the potential to gain a deeper understanding of the emitter’s structure \cite{tawfik_first-principles_2017}.
Resonant excitation reveals linewidths within the FL of  $(55 \pm 10)$ MHz. 
The spectral stability of the QEs enabled us to record two-dimensional intensity maps of the emitter under resonant excitation without any decrease in fluorescence due to blinking or spectral drifts.
To achieve the observed spectral stability we employed annealing techniques resulting in a significantly decreased inhomogeneous linewidth recorded in Photoluminescence spectroscopy (PL) from on average $\approx 340$ GHz to on average $\approx 213$ GHz statistically evaluated on 121 QEs.
Resonant excitation further reduces the inhomogeneous linewidth from 293 GHz in PL to 67 GHz in PLE specified for one specific QE.
Investigations of the phonon sideband indicates a phonon mode at 45 THz, in accordance with reference \cite{kern_ab_1999,tohei_debye_2006,reich_resonant_2005,sanchez-portal_vibrational_2002,yu_ab_2003}, that we utilize as detection channel for resonant excitation.
Together with dipole-like, linear polarization with high contrast in emission our results put hBN on the forefront of single photon sources with FL photon emission.
Our investigated platform consists of two-dimensional hBN flakes mounted on a silver mirror which enables direct integration into photonic structures such as optical resonators, for example, for spin-photon interfaces with Purcell-enhanced, lifetime-limited, single-photon emission.

\begin{figure}[htbp!]
\includegraphics[width=0.5\textwidth]{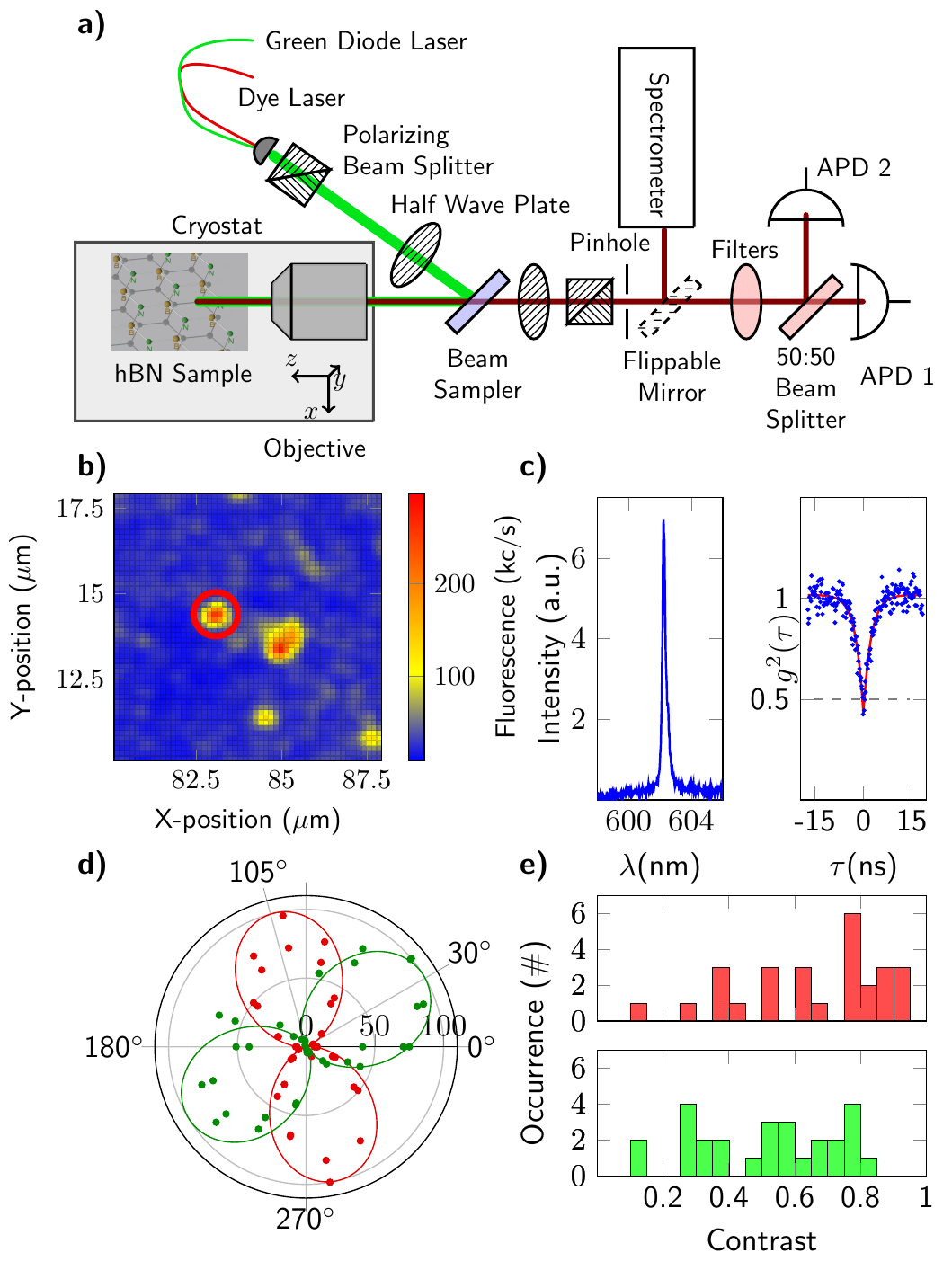}
\caption{\label{fig:fig1} (a) Experimental Setup. The hBN sample is cooled to liquid helium temperatures of $\approx 5$ K in a continuous flow cryostat and investigated via a custom-built confocal microscope with an high NA Objective of 0.9. 
We perform PL and polarization measurements in off-resonant excitation with a 532 nm green diode laser and with a power of up to 250 $\mu$W in front of the objective. 
A spectrometer with a maximal resolution of 16 GHz records the PL spectra and a HBT setup enables to extract second-order, photon correlations. 
For Polarization measurements we use additional polariser and a $\lambda$ -half wave plate.
Resonant excitation was performed using a CW Matisse DS Dye Laser.
(b) Confocal scan (max. $200 \times 200$ $\mu$m) of the emitters in hBN flakes with off-resonant excitation.
(c) Spectrum and second-order correlation of the emitter marked in b). 
$g^2(0)=0.42$ reveals single photon character of the QE.
Background subtraction results in $g^2(0)=0.01$.
The lifetime is then extrapolated to ($2.37\pm0.12$) ns corresponding to a natural linewidth of $( 67.2 \pm 3.2 )$ MHz.
(d) and (e) We find emission and excitation are missaligned by 75$^{\circ}$.
The  polarization contrast of emission (red data) has an average of 0.6 and for excitation (green data) an average of 0.5.}
\end{figure}

QEs are prepared from solvent-exfoliated hBN flakes purchased from Graphene Supermarket. 
The flakes are dispersed on a marked silicon substrate coated with a 300 nm thermal oxide capping layer. 
Activation of the emitters involves annealing at 850 $^\circ$C for 30 minutes under a fluxing Ar atmosphere, at 1 torr. 
For the transfer process onto a silver mirror, the as-prepared silicon substrates with hBN flakes were allowed to float on the surface of a 2 M NaOH solution in a glass container. 
The container was then heated to 90 $^\circ$C for 2 h to accelerate the silicon etching process. 
After such a period, the PMMA-coated hBN flakes were completely detached from the substrate and floating on the surface of the alkaline solution.
The membranes were repeatedly rinsed in deionized water for four times, loaded onto the silver mirror, and dried at 40 $^\circ$C on a hot plate. 
The substrates were then post-baked at 120 $^\circ$C for 20 min to increase the adhesion between the hBN flakes and the mirror surface. 
The PMMA film was subsequently removed with warm acetone, leaving behind only the hBN flakes on the silver mirror surface.
The hBN flakes inheriting color center through the synthesis process \cite{li_atomically_2016} are cooled in a continuous-flow cryostat to liquid helium temperature, see Fig. \ref{fig:fig1}(a), and scanned in a confocal microscope Fig. \ref{fig:fig1}(b).
The excitation and emission polarization is extracted by a linear polarizer and a $\lambda$-half wave plate probing the incident or emitted polarization via spectrometer and/or via APD counts. 
\\
For resonant excitation we use a CW Matisse DS Dye laser system, which has a mode hop-free scanning range of \textgreater 40 GHz and a line-width of \textless 250 kHz.
The wavelength of the laser is monitored via a wave meter from High Finesse with 500 kHz resolution and acquisition frequency up to 600 Hz. 
Photo Luminescence Excitation (PLE) is measured by detecting the phonon sideband emission while the direct laserlight is blocked with a LP Filter.

\begin{figure}
\includegraphics[width=0.5\textwidth]{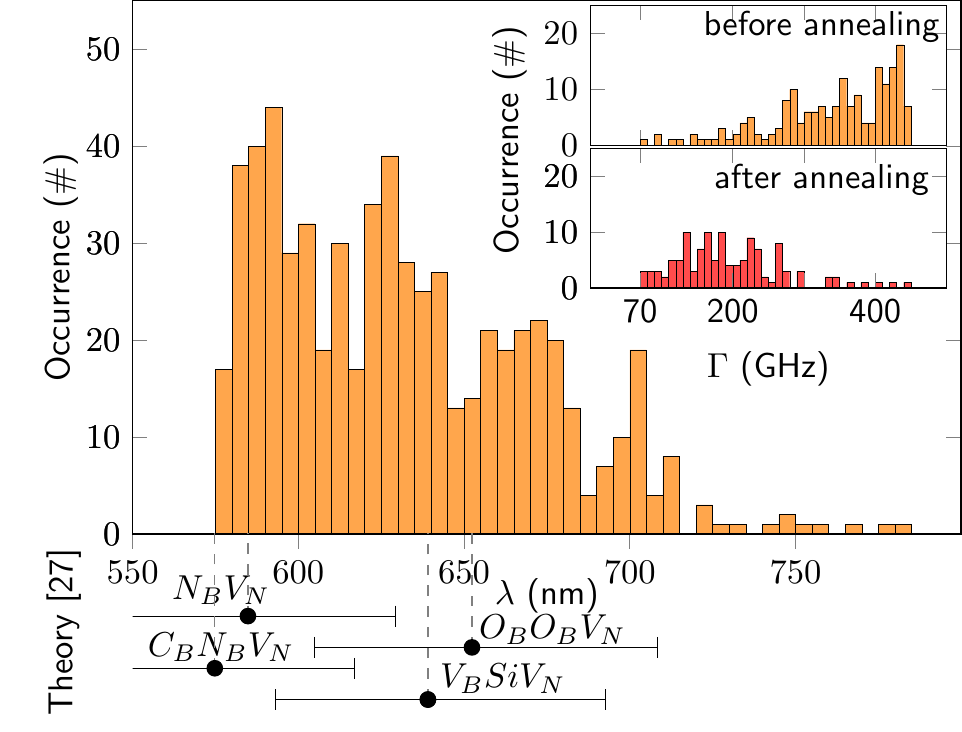}
\caption{\label{fig:fig2} Histogram of 627 individual ZPLs binned within 10 nm cluster around four emission wavelengths at 590 nm,  630 nm, 670 nm and 700 nm. Another cluster is indicated at 750 nm. We  compare this distribution to DFT calculated ZPL energies  of color center  with an error-margin of 0.3 eV [S. A. Tawfik, Nanoscale \textbf{9}, (2016)]. Inset: 627 measured linewidths before (yellow bars) and 121 linewidths after annealing (red bars) in vacuum at 500$^{\circ}$C for 1 h. The average linewidth decreases by about 120 GHz.}
\end{figure}

We evaluate emission spectra from many different QEs and measure second-order autocorrelation function in a Hanbury-Brown and Twiss setup (HBT) with $g^2(\tau=0)< 0.5$ proofing the single photon character of the QE, see Fig. \ref{fig:fig1}(c).
The data fit is based on $g^2(\tau)$ model for two level system
\begin{equation}
g^2(\tau)=1-a\cdot e^{|\tau|/\tau_0}.
\end{equation}
Various sites have lifetimes in the range of $\tau_0=$ (1.53 - 2.88) ns resulting in natural linewidth of $\Gamma=$ (55 - 84) MHz, in agreement to previous results \cite{tran_robust_2016,sontheimer_photodynamics_2017,tran_resonant_2017}.\\
The polarization measurements show high contrast for the emission process as high as 0.9 and with an average of 0.6, see Fig. \ref{fig:fig1}(d) and (e).
The contrast is defined as :

\begin{equation}
C=\frac{I_{max}-I_{min}}{I_{max}+I_{min}}
\end{equation}
Deviation from perfect polarization contrast may originate from non-perfect horizontal alignment of the hBN flakes with respect to the substrate.
The polarization in off-resonant excitation results in a reduced contrast of on average 0.5, indicating that polarization is not preserved in phonon-assisted excitation.
\\
PL spectra are recorded for 627 individual ZPL lines, distributing from 580 nm to 800 nm, see Fig. \ref{fig:fig2}(a).
The ZPL emission frequencies cluster in four distinct positions, with peaks at 590 nm, 630 nm, 670 nm and 700 nm and with an overall decreasing occurrence with increasing wavelength. Another cluster is indicated at 750 nm.
Each cluster could originate from a specific emitter composition with an inhomogeneous spectral linewidth of about 50 nm.
In-depth knowledge on the QEs formation could be gained by comparison with ab-inito calculations as developed in references \cite{tawfik_first-principles_2017,abdi_spin-mechanical_2017}. 
The grouping into four distinct regions could be correlated to the simulated compositions of the defects as depicted in Fig. \ref{fig:fig2} bottom.
The high energy emission is likely to correlate to the neutrally charged $N_BV_N$ defect \cite{tawfik_first-principles_2017,abdi_spin-mechanical_2017}.
The lower energy emission is likely to correlate to the proposed carbon related defects, the positively charged $V_NC_B$ \cite{abdi_color_2017} and the $C_NV_B$ defect \cite{tawfik_first-principles_2017}. 
While our work is beyond the scope of concretely isolating the selected defect, we note that these defects have similar crystallographic symmetry and therefore are expected to yield similar resonant excitation behavior.
Our data could be used to refine and calibrate future ab-inito calculations. 
The observed spectral diffusion is in accord with the calculated structure that suggests a persistent dipole moment.
\\
Before annealing the sample in vacuum at 500$^{\circ}$ C over 1h the average linewidth of an ensemble of 612 sites is 339 GHz, see upper inset Fig. \ref{fig:fig2}. 
After annealing, an ensemble of 121 individual sites shows a decreased linewidth of on average 213 GHz, see lower inset Fig. \ref{fig:fig2}.
The improved linewidth can be explained by elimination of chemical residue on the surface which also improves the broad background fluorescence as reported in reference \cite{garcia_effective_2012}.
However, in both cases the narrowest observed linewidth of $\approx 70$ GHz is still limited by significant spectral diffusion, which could arise from charge fluctuations on the surface \cite{wolters_measurement_2013,pelton_modified_2015}.

\begin{figure}[htbp!]
\includegraphics[width=0.5\textwidth]{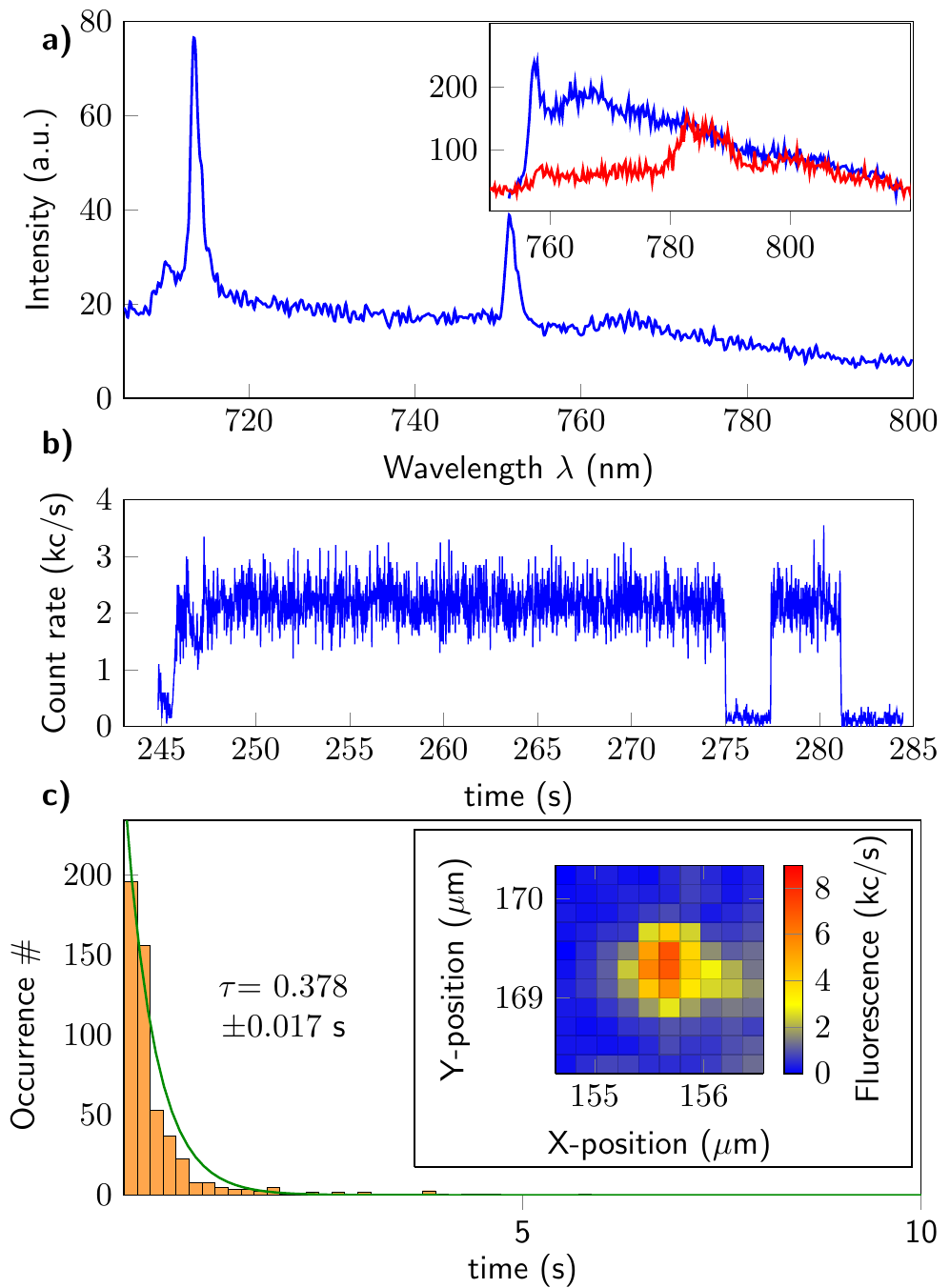}
\caption{\label{fig:fig3} (a) PL spectrum of the investigated emitter at emission wavelength of 713 nm with an additional emitter at 755 nm. Inset: The sideband is measured with a 750 LP filter with off-resonant (blue data) and resonant (red data) excitation. Resonant excitation separates the phonon sideband of the 713 nm emitter from fluorescence of the additional emitter at 755 nm. The phonon mode at 785 nm with 185 meV (45 THz, 1500 cm$^{-1}$) matches the reported local LO mode of hBN \cite{kern_ab_1999,tohei_debye_2006,reich_resonant_2005,sanchez-portal_vibrational_2002}.  (b) Fluorescence time traces are stable under resonant excitation for as long as 30 seconds. (c) Statistics of the on-time of the fluorescence traces discloses a time constant of 0.37 s for spectral instabilities. Inset: The spectral stability is long enough to extract two-dimensional maps of the emitter in resonant excitation without significant fluctuations of the fluorescence signal.}
\end{figure}

Resonant excitation further reduces the linewidth with respect to off-resonant excitation. 
We investigate resonant excitation of a QE with an emission line at 713 nm, which showed an inhomogeneous linewidth of 293 GHz, see Fig. \ref{fig:fig3}(a), and which is well-suited for our resonant excitation laser. 
The emission spectrum indicates a second emitter at $\approx 755$ nm emission wavelength as well as phonon-assisted emission at 785 nm that we use for the detection during resonant excitation, while blocking the laser with a 750 nm LP filter.
To separate the phonon sideband (PSB) emission from our emitter at 713 nm from fluorescence originating from the additional emitter at 755 nm we record the spectrum  in off-resonant excitation (blue spectra, inset  Fig. \ref{fig:fig3}(a)), as well as in resonant excitation (red spectra, inset  Fig. \ref{fig:fig3}(a)).
The recorded phonon peak at around  785 nm corresponds to a phonon mode of hBN at 187 meV (45 THz or 1500 $cm^{-1}$) as reported in references \cite{kern_ab_1999,tohei_debye_2006,reich_resonant_2005,sanchez-portal_vibrational_2002}.
According to reference \cite{yu_ab_2003} this mode is assigned to be a longitudinal optical mode with a group symmetry of $E_{1u}$.
\\
Resonant excitation reveals blinking in the fluorescence signal as shown in Fig. \ref{fig:fig3}(b), with an 'OFF' state corresponding to a background level of 300 c/s, a clear 'ON' state corresponding to $(3000\pm 500)$ c/s and no visible photo bleaching.
Eventually very long periods of stable fluorescence occur for many seconds which we attribute to slowly fluctuating, trapped carrier-induced Stark shifts followed by large spectral jumps as reported in \cite{shotan_photoinduced_2016}.
Such periods of stable fluorescence enable us to perform resonant confocal scans without any significant change in the fluorescence signal, see inset Fig. \ref{fig:fig3}(c).
Detailed statistics of the blinking timescale is unfolded by a histogram of the 'ON' times over their occurrence with a binning of 0.2 s in Fig. \ref{fig:fig3}(c).
A fit with an exponential decay reveals a single time constant of $(\tau=0.378 \pm0.017)$ s.
\\

\begin{figure}[htbp]
\includegraphics[width=0.5\textwidth]{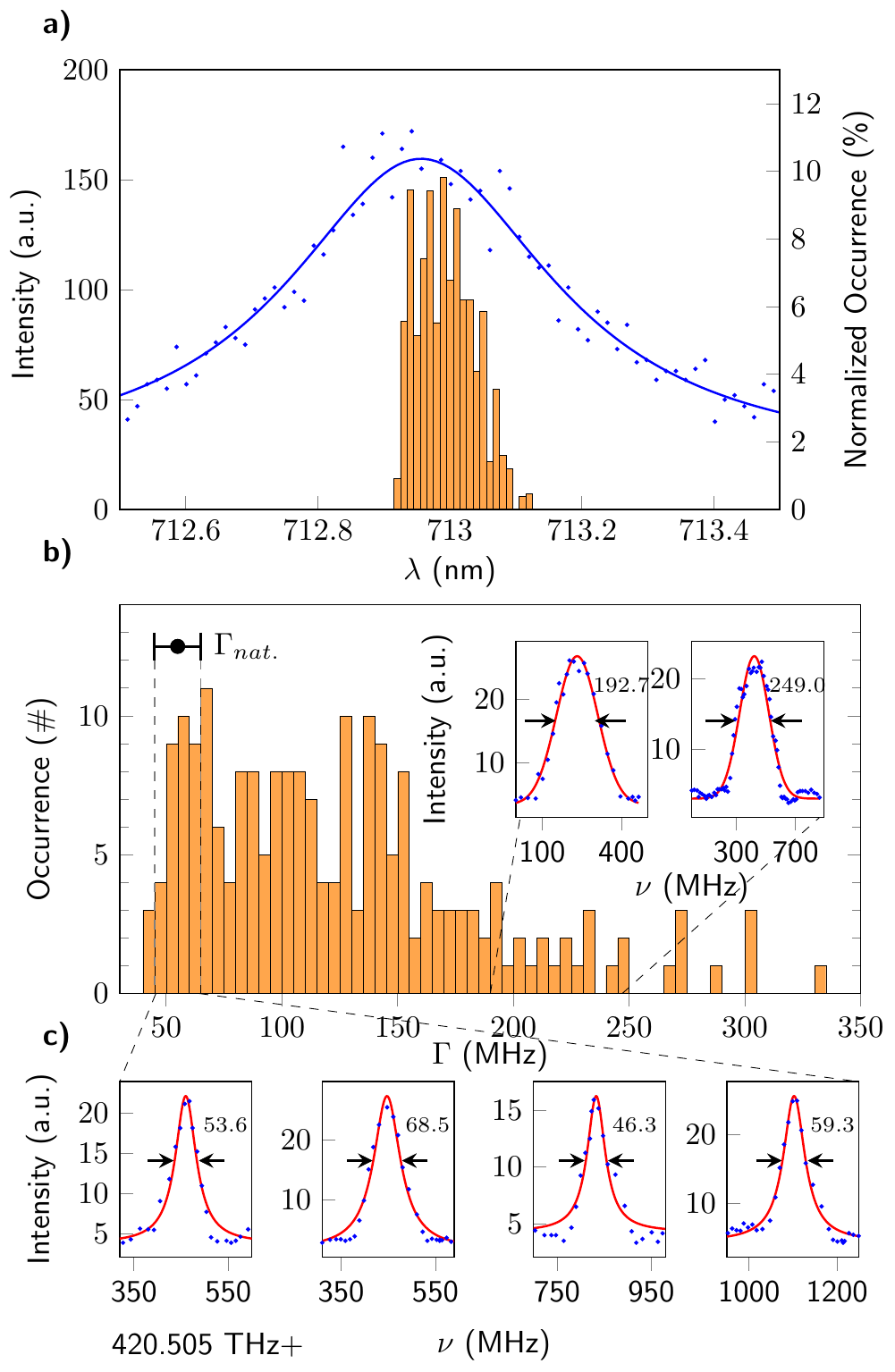}
\caption{\label{fig:fig4} (a) The PL spectrum (blue data, red Gaussian fit) has an inhomogeneous linewidth of $(293.39 \pm 8.13)$ GHz. The histogram of 207 PLE lines measured for the same emitter over a range of 200 GHz reveals an inhomogeneous linewidth of $(67.5 \pm 9.5)$ GHz. (b) The homogeneous linewidth extracted from more than 207 PLE linescans of the same emitter unfolds linewidths of $(124.5 \pm 60.5)$ MHz. The large variance hints at ongoing dynamics on the second timescale of the performed scans. (c) About 42 lines of that specific emitter  lie within the error margin of the natural linewidth of $(55 \pm 10)$ MHz.}
\end{figure}

In order to record linewidth in resonant excitation we perform PLE line scans faster than characteristic diffusion time $\tau=0.378$ s.
The slowest applied scanning speed of the resonant laser is about 133 MHz/s with a recording frequency of 50 Hz, enable resonant line scans faster than the diffusion time. 
We obtained the best signal to noise ratio for 4-6 $\mu$W excitation power in front of the objective, while we made sure to be well below saturation. 
\\
Several PLE scans performed over range of 200 GHz of the same QE emitting at 713 nm uncover the histogram displayed in Fig. \ref{fig:fig4}(a).
Scans beyond this 200 GHz revealed no PLE signal.
We normalize the statistics of occurrences in Fig. \ref{fig:fig4}(a) (right axis), taking into account the number of scans per bin (6 GHz).
The normalized PLE statistic discloses a spectral width of $( 67.5 \pm 9.5)$ GHz much narrower than the inhomogeneous PL linewidth of $(293.39 \pm 8.13)$ GHz (blue data and Gaussian fit in Fig. \ref{fig:fig4}(a)).
We filtered out all data which showed blinking during laser scanning.
The lifetime of this line is $\approx 2.88$ ns giving a FL linewidth of $ \Gamma_{nat.}= (55.26 \pm 0.19)$ MHz.
While the homogeneous linewidth of 124.5 MHz is still more than twice as large as the FL linewidth the large variance of 60.5 MHz after averaging over all 207 lines indicates on-going dynamics on the timescale of the resonant laser scan.  
Details of the linewidth statistics in Fig. \ref{fig:fig4}(b) shows a vast spread with linewidth up to 200 MHz.
Two examples of the broadened resonant lines are shown in the inset.
Note that these lines are fit well with Gaussian function, confirming that the predominant mechanism for the broadening is spectral diffusion.
Nevertheless, about 42 lines out of all 207 lines are within the error margin of 10 MHz of the natural linewidth.
We fitted the lines with a Lorentz function and four example lines are shown in Fig. \ref{fig:fig4}(c), displaying FL lines with linewidth of $(60 \pm 10)$ MHz - $(46 \pm 10)$ MHz.
The uncertainty of 10 MHz arises from the systematic error of the measurement and the evaluation algorithm.  

In summary, we have demonstrated FL, single-photon emission from defect centers in hBN without significant spectral diffusion or spectral instability for as long as 30 seconds.
Our results render possible photon interference experiments at high rates.
Conservative extrapolation from the reported count rates of $\approx 2000$ counts/sec, measured well below saturation and detecting only small fraction of the sideband emission, yields indistinguishable single photon emission rates much larger than $20 000$ counts/sec in the current setting assuming a Debye-Waller factor of 0.82 as reported in \cite{tran_quantum_2016_nature}.
\\
The investigated platform consists of hBN flakes placed on a silver mirror and opens new perspective for resonance fluorescence experiments or investigations of mirror image interactions. 
The architecture demonstrates the compatibility with photonic platforms, in particular optical resonators, for Purcell-enhanced single photon emission  facilitating indistinguishable photon rates up to GHz rates.
Some QE in hBN could potentially inhere long spin coherence time \cite{abdi_spin-mechanical_2017}. 
In this context, our work puts hBN on the forefront of potential candidates for realizing Quantum repeaters or Quantum networks based on a compact solid state system with photonic channels connecting remote spin systems.
In addition, our detailed investigations of more than 627 QEs in hBN give new input for theoretical predictions of the defects composition by comparing ab-initio simulations \cite{abdi_color_2017}  with the reported emission frequencies.\\
\begin{acknowledgments}
AK acknowledges the generous support of the DFG, the Carl-Zeiss Foundation, IQST, the Wissenschaflter-R\"uckkehrprogramm GSO/GZS. 
FJ acknowledges support of the DFG,BMBF, VW Stiftung and EU (ERC, DIADEMS).
I.A. acknowledges the generous support of the Alexander van Humboldt foundation, and the Asian Office of Aerospace Research \& Development (grant \# FA2386-17-1-4064). 
Experiments performed for this work were operated using the Qudi software suite \cite{binder_qudi:_2017}.

\end{acknowledgments}

\appendix

\newpage

\nocite{*}
\bibliography{Narrowband_emitters_FL_hBN_A_Dietrich_et_al}

\end{document}